%% file: ksks.tex
\documentclass[preprint,showpacs,superscriptaddress,preprintnumbers,amsmath,amssymb]{elsart}

\usepackage{graphicx}
\usepackage{dcolumn}
\usepackage{bm}

\begin{document}
\begin{frontmatter}
\title{A Study of $D^0\rightarrow K^0_SK^0_SX$ Decay Channels}

\date{\today}

\input{author.tex}

\begin{abstract}

Using data from the FOCUS experiment (FNAL-E831), we report on  
the decay of $D^0$ mesons into final states containing more than one $K^0_S$. We
present evidence for two Cabibbo favored decay modes, $D^0\rightarrow
K^0_SK^0_S K^- \pi^+$ and $D^0\rightarrow K^0_SK^0_S K^+ \pi^-$, and measure
their combined branching fraction relative to $D^0\rightarrow \overline{K}\,\!^0\pi^+\pi^-$
to be 
$\frac{\Gamma(D^0\rightarrow K^0_SK^0_SK^{\pm}\pi^{\mp})}{\Gamma(D^0\rightarrow
\overline{K}\,\!^0\pi^+\pi^-)}$ =
0.0106 $\pm$ 0.0019 $\pm$ 0.0010. Further, we report new measurements of
$\frac{\Gamma(D^0\rightarrow K^0_SK^0_SK^0_S)}{\Gamma(D^0\rightarrow
\overline{K}\,\!^0\pi^+\pi^-)}$ = 0.0179 $\pm$ 0.0027 $\pm$ 0.0026,
$\frac{\Gamma(D^0\rightarrow K^0\overline{K}\,\!^0)}{\Gamma(D^0\rightarrow
\overline{K}\,\!^0\pi^+\pi^-)}$ = 0.0144 $\pm$ 0.0032 $\pm$ 0.0016, and 
$\frac{\Gamma(D^0\rightarrow K^0_SK^0_S\pi^+\pi^-)}{\Gamma(D^0\rightarrow
\overline{K}\,\!^0\pi^+\pi^-)}$ = 0.0208 $\pm$ 0.0035 $\pm$ 0.0021 where the first error is statistical and the second
 is systematic.

\end{abstract}
\end{frontmatter}

\section{\label{sec:level1}Introduction}

Detailed measurements of rare exclusive decay modes of charm mesons provide
a powerful way to probe the details of charm decay such as the contributions of $W$ exchange
diagrams and final state interactions.  These measurements aid our understanding of the interplay
between the weak and strong interactions, mainly for multibody decays where theoretical predictions are poorer than for two body decays. 
 We report the first observation
of $D^0\rightarrow
K^0_SK^0_S\pi^{\pm}K^{\mp}$, and new measurements of $D^0\rightarrow K^0_SK^0_SK^0_S$, 
$D^0\rightarrow K^0_SK^0_S$,   and
$D^0\rightarrow K^0_SK^0_S\pi^+\pi^-$.  

The fixed-target charm photoproduction experiment FOCUS, an upgraded version of E687 \cite{nim.320.519}, collected 
data during the 1996--1997 fixed-target run at
Fermilab. A photon beam is derived from the bremsstrahlung of secondary electrons and positrons produced from the 800
GeV/$c$ Tevatron proton beam. The photon beam interacts with a segmented beryllium-oxide target. The average photon
energy for reconstructed charm events is 180 GeV. \\
Two silicon microvertex systems provide excellent separation between the production
and charm decay vertices. One silicon strip detector called target silicon (TS)
is embedded in the BeO target segments \cite{TSNIM}. The other silicon strip detector (SSD) is located downstream 
of the target region. Charged particles are tracked and momentum analyzed with five
stations of multiwire proportional chambers in a two magnet forward spectrometer. Three
multicell threshold \v{C}erenkov detectors are implemented to identify electrons, pions, kaons,
and protons \cite{ceren.0108011}.

\section{\label{sec:level2}Reconstruction of individual particles: $K^0_S$, $K$, and $\pi$}

A detailed description of individual $K^0_S$ reconstruction in the FOCUS
spectrometer can be found elsewhere \cite{vees.0109028}. Briefly, $K^0_S$'s
are identified in several different regions depending on the $K^0_S$ decay length; 
upstream of the magnet with silicon strip information, upstream of the magnet
without silicon strip information, and inside of the magnet. Depending on 
whether the daughter pions from the $K^0_S$ pass through the second magnet, 
and hence have a well defined momentum, a series of different techniques are employed.
Roughly 15\% of the  $K^0_S$ are found upstream of the silicon system and can be used
in locating the $D^0$ decay vertex.  When there are two or more $K^0_S$ in a decay 
channel, the daughter tracks must all be distinct and can not be shared between
$K^0_S$ candidates. The reconstructed mass of the $K^0_S$ must be within three standard deviations of the nominal
$K^0_S$ mass.

All charged tracks from the charm decay must be singly linked to the silicon microstrip, be of good quality,
and inconsistent with zero degree photon conversion. 
\v{C}erenkov particle identification (PID) for charged particles is performed by constructing a
log likelihood value ${\mathcal W}_i$ for the particle
hypotheses ($ i = e, \pi, K, p$).
The $\pi$ consistency of a track is defined by $\Delta {\mathcal W}_\pi = {\mathcal W}_{min} - {\mathcal
W}_\pi$,
where ${\mathcal W}_{min}$ is the minimum ${\mathcal W}$ value of
the other three hypotheses. Similarly, we define $\Delta {\mathcal W}_{K,\pi} = {\mathcal W}_\pi -
{\mathcal W}_K$ and $\Delta {\mathcal W}_{K,p} = {\mathcal W}_p -
{\mathcal W}_K$
for kaon identification and we
require $\Delta {\mathcal W}_{K,\pi} > 1 $ and $\Delta {\mathcal W}_{K,p} > -2 $ for kaons
and $\Delta {\mathcal W}_\pi > -5$ for pions.

\section{\label{sec:level3}Reconstruction of $D^0$ candidates}
The $D^0$ candidates are formed by making a vertex hypothesis for the daughter 
particles whenever possible.
We use the $K^0_S$ candidates and two charged tracks of the correct charge combination 
to form a $D^0$ candidate.
The confidence level of the decay vertex of the $D^0$ candidate is 
required to be greater than 1\%.  The combined momentum vector located at the 
decay vertex forms the $D^0$ track. Using the $D^0$ track as a seed track
for  the candidate driven vertexing algorithm \cite{nim.320.519}, we search 
for a production vertex in
which the confidence level is greater than 1\% and
the primary multiplicity including the seed track must have at least 3 tracks. 
The production 
vertex must be inside the target.
The significance of separation between the production and the decay vertices 
($L/\sigma_L$) must be greater than 4  for $D^0\rightarrow
K^0_SK^0_S\pi^{\mp}K^{\pm}$ and greater than 6 for $D^0\rightarrow
K^0_SK^0_S\pi^+\pi^-$.
Different values are chosen for the $L/\sigma_L$ cut due to the level of background for the
two channels, but scans are performed in  $L/\sigma_L$ as part of the systematic uncertainty
studies.

A stand-alone vertex algorithm is used to reconstruct the primary vertex of the event for decay modes without charged
tracks \cite{nim.320.519}.
All the silicon microvertex tracks in the event, excluding those
already assigned to pions used to reconstruct $K^0 _S$, are used to form all possible
vertices of the event with a confidence level greater than 1\%: we then choose
the primary vertex to be the highest multiplicity vertex. Ties are resolved
choosing the most upstream vertex.

The signal channels are normalized to $D^0\rightarrow \overline{K}\,\!^0\pi^+\pi^-$ which is 
the most abundant $D^0$ decay mode containing a $K^0_S$.
We use only the most basic cuts in this analysis. The invariant mass distribution for $D^0\rightarrow
K^0_S\pi^+\pi^-$ at an $L/\sigma_L > 4$ is presented in  Fig.~\ref{figone:mass}a.  To determine
the reconstruction efficiency of this channel we use a Monte Carlo simulation using our best
determined decay distribution to generate $D^0\rightarrow
\overline{K}\,\!^0\pi^+\pi^-$. The reconstructed Monte Carlo sample was six times larger than the data sample.
 We fit our signals from both data and Monte Carlo 
with a Gaussian for the signal and a $2^{nd}$ degree polynomial for the background using a maximum
likelihood fit.  As our most copious $K^0_S$ decay channel, we use the $D^0\rightarrow K^0_S\pi^+\pi^-$
mode to search for variations versus reconstructed $K^0_S$ categories. We identify 
a systematic uncertainty on $K^0_S$ reconstruction of 7.1\%. 

The systematic uncertainty for each branching fraction is independently determined.
Several tests are performed. To check the stability of the branching fractions we fix the
cuts to their standard values and vary one cut at a time. Cuts which are scanned include
particle identification, significance of separation, isolation of the secondary vertex
from any other track, and confidence level of the secondary vertex. We also perform split
sample systematics by dividing our samples into two sub-samples based on momentum and 
data--taking period.  We study fit variant systematics by calculating the branching fraction
when the signals are fit with $1^{\mathrm{st}}$, $2^{\mathrm{nd}}$, or $3^{\mathrm{rd}}$ degree polynomials. 

We look for Monte Carlo resonance systematics by searching for a difference in reconstruction
efficiency between non-resonant generation and a specific resonance channel. 
We also include a systematic uncertainty
for the absolute tracking efficiency (0.2\% per difference in numbers of tracks) and 
for $K_S^0$ reconstruction efficiency (7.1\% per $K_S^0$).  Finally, we include a small contribution
to the systematic uncertainty due to Monte Carlo statistics. All systematic uncertainties
are added in quadrature.

\section{\label{sec:level4}\bf The $D^0\rightarrow K^0_SK^0_SK^{\pm}\pi^{\mp}$ Decay Mode}

\begin{figure}[h!]
  \begin{center}
     \includegraphics[width=4.55cm,height=6.8cm]{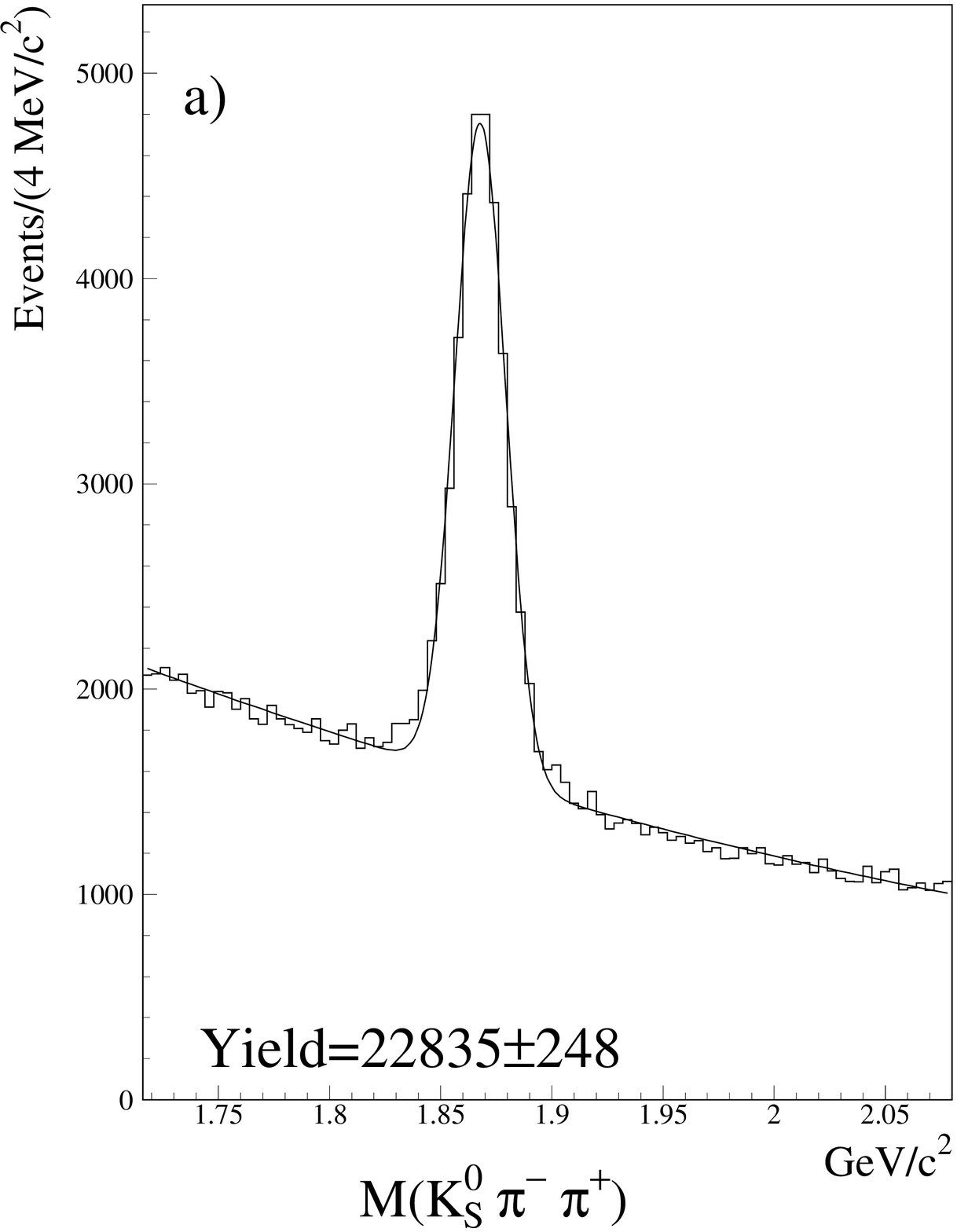}
     \includegraphics[width=4.55cm,height=6.8cm]{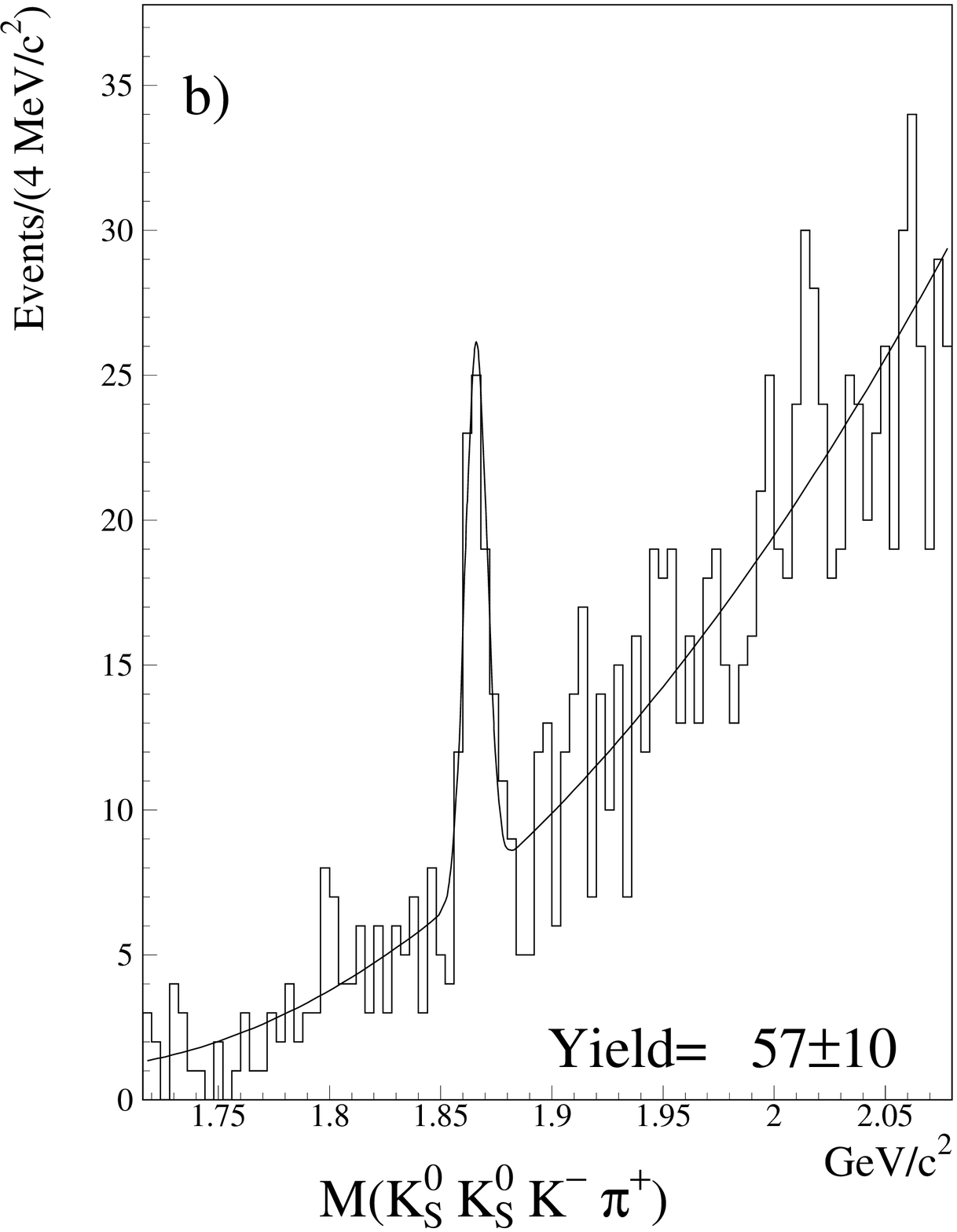}
     \includegraphics[width=4.55cm,height=6.8cm]{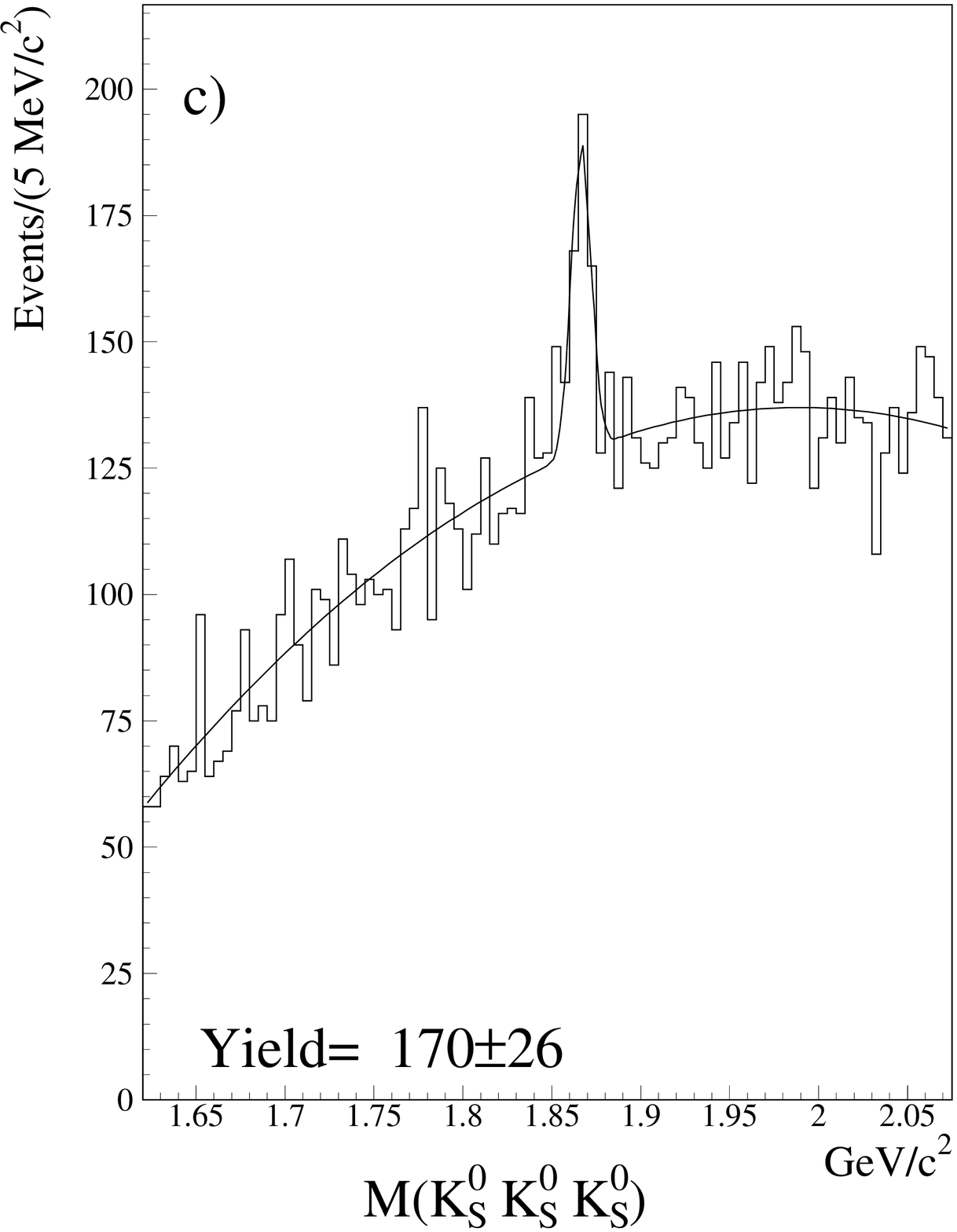}
  \caption{ Invariant mass distribution for various $D^0$ final states: 
    (a) Reconstructed mass of $D^0 \rightarrow K^0_S\pi^+\pi^-$. There are $22835 \pm 248$ events 
with a sigma of 11.4 MeV$/c^2$.  
    (b) Reconstructed mass of $D^0 \rightarrow K^0_SK^0_SK^{\pm}\pi^{\mp}$. There are $57 \pm 10$ events  
with a sigma of 5.0 MeV$/c^2$.  
    (c) Reconstructed mass of $D^0 \rightarrow K^0_SK^0_SK^0_S$. There are $170 \pm 26$ events 
with a sigma of 5.6 MeV$/c^2$. 
   }
  \label{figone:mass}
\end{center}
\end{figure}

The $D^0$ final state $K^0_SK^0_SK^{\pm}\pi^{\mp}$ has not been
observed by previous experiments. While the $D^0$ decay modes $D^0\rightarrow
{\overline{K}\,\!^0}{\overline{K}\,\!^0}K^+\pi^-$,
 $D^0\rightarrow \overline{K}\,\!^0K^0K^-\pi^+$, and
$D^0\rightarrow \overline{K}\,\!^0K^0K^+\pi^-$ can result
in a final state  $K^0_SK^0_SK^{\pm}\pi^{\mp}$, the decay mode
$D^0\rightarrow \overline{K}\,\!^0K^0K^+\pi^-$ is doubly Cabibbo suppressed
and we assume that this mode is small compared to the
other two Cabibbo favored modes.
To distinguish the two Cabibbo favored decays we use the
charge of the soft pion in the decay sequence $D^{*+} \rightarrow D^0\pi^+$.
First, we demonstrate a signal for the two channels combined without
using a soft pion tag and then we argue that we have evidence for independent observations in both channels.
In Fig.~\ref{figone:mass}b we show the invariant mass plot for $D^0\rightarrow
K^0_SK^0_SK^{\pm}\pi^{\mp}$ at $L/\sigma >$~4.
We fit the plot
with a Gaussian for the signal and a $2^{nd}$ degree polynomial for the
background
using a maximum likelihood fit and we find  $57 \pm 10$ events.

We find no evidence for resonant substructure in $D^0 \rightarrow
K_S^0K_S^0K^\pm\pi^\mp$ and use a non-resonant Monte Carlo simulation to
obtain the efficiency.  As a systematic check we find a 2.6\% difference
in efficiency from a $D^0 \rightarrow a_0(980)^+K^*(892)^-$ Monte Carlo
simulation.
                                                                                
Correcting for efficiency, we find:
\begin{equation}
 \frac{\Gamma (D^0 \rightarrow K^0_SK^0_SK^{\pm}\pi^{\mp})}
 {\Gamma (D^0 \rightarrow \overline{K}\,\!^0\pi^+\pi^-)} = 0.0106 \pm 0.0019 \pm 0.0010.
\end{equation}
We have included systematic uncertainty contributions from split samples (0), fit variations (0.0007),
possible resonant decay (0.0001), Monte Carlo statistics (0.0001), absolute tracking efficiency (0.0001), and 
variation between $K^0_S$ reconstruction categories (0.00075).

\begin{figure}[th!]
\begin{center}
\includegraphics[width=4.55cm,height=6.8cm]{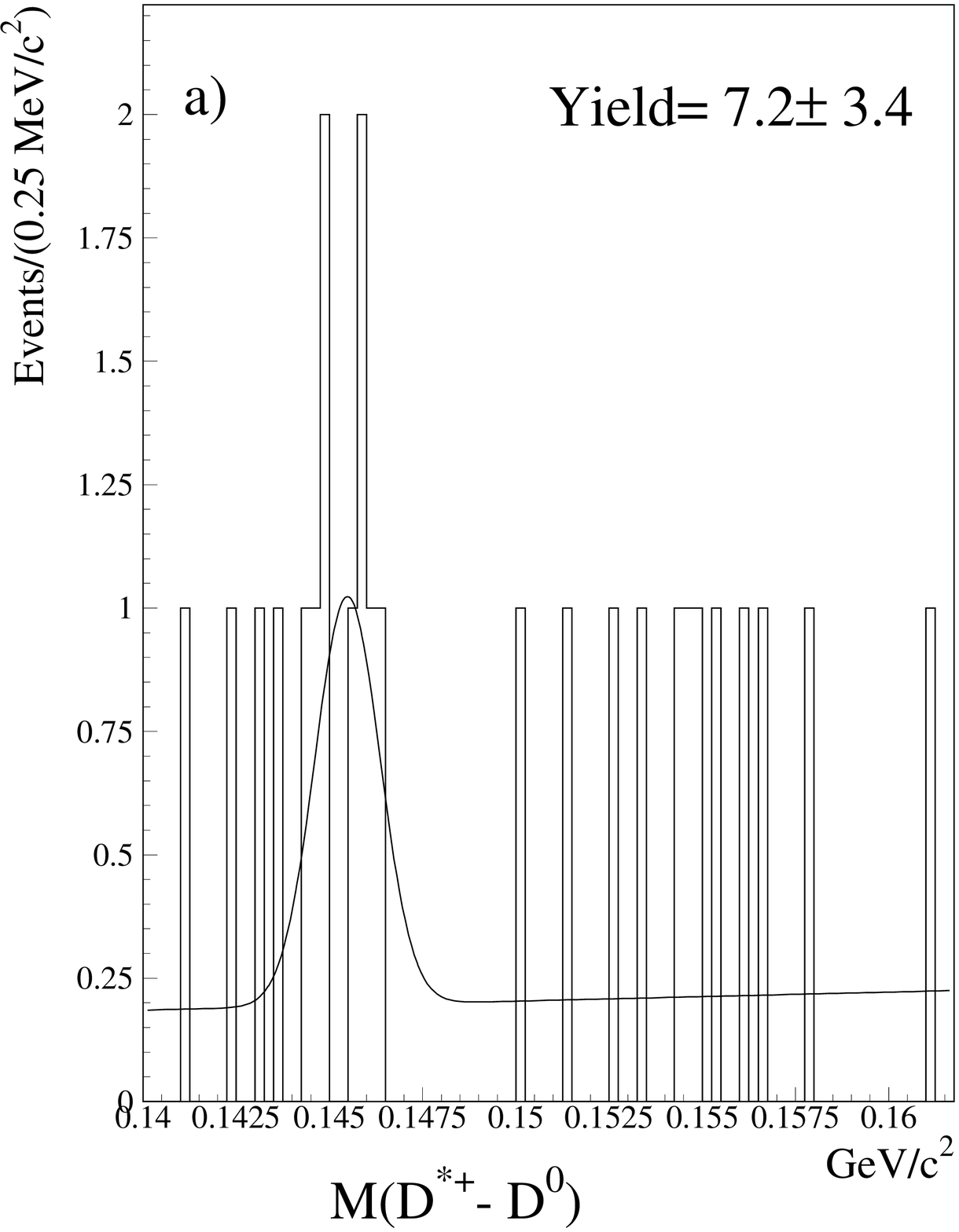}
\includegraphics[width=4.55cm,height=6.8cm]{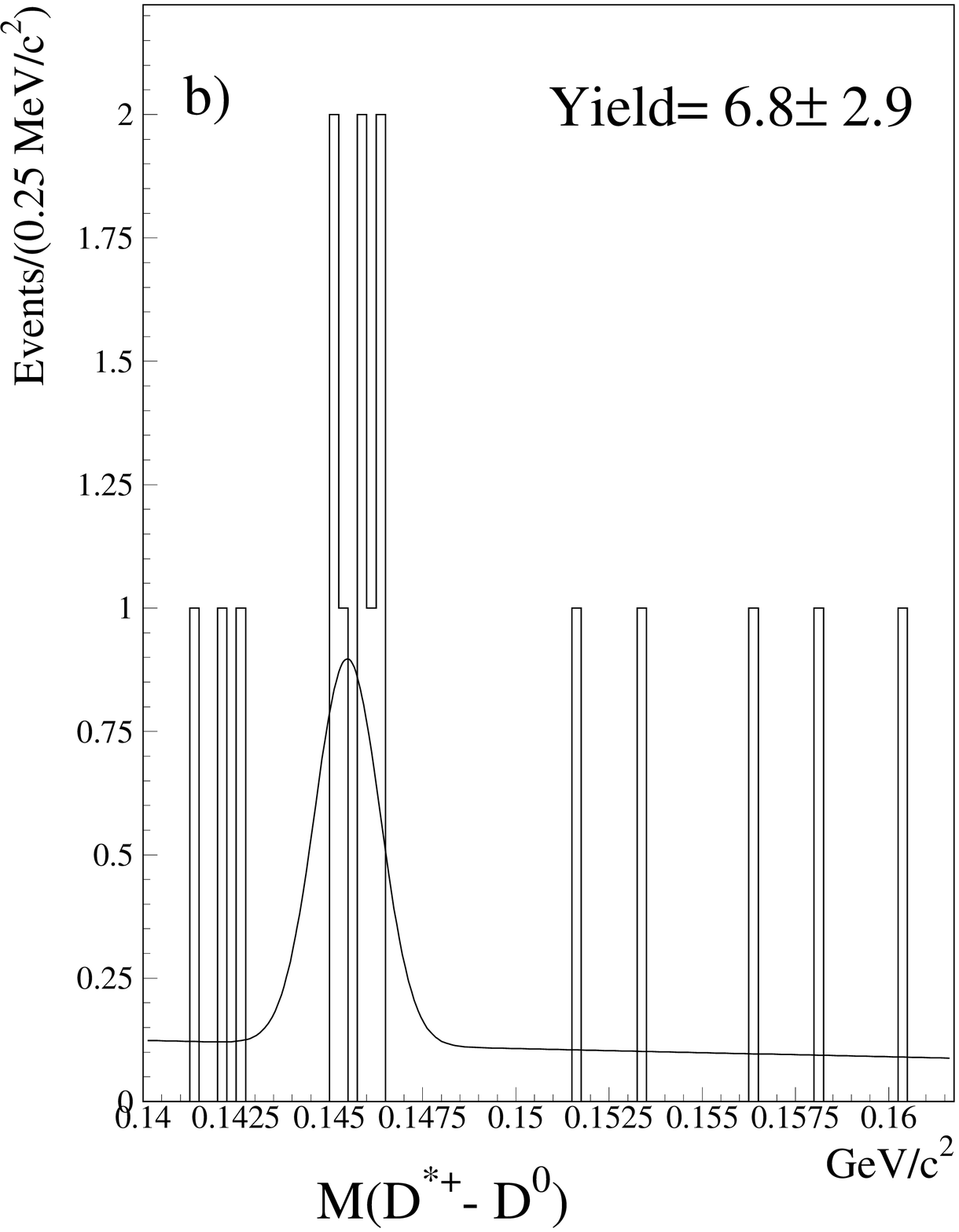}
\includegraphics[width=4.55cm, height=6.8cm]{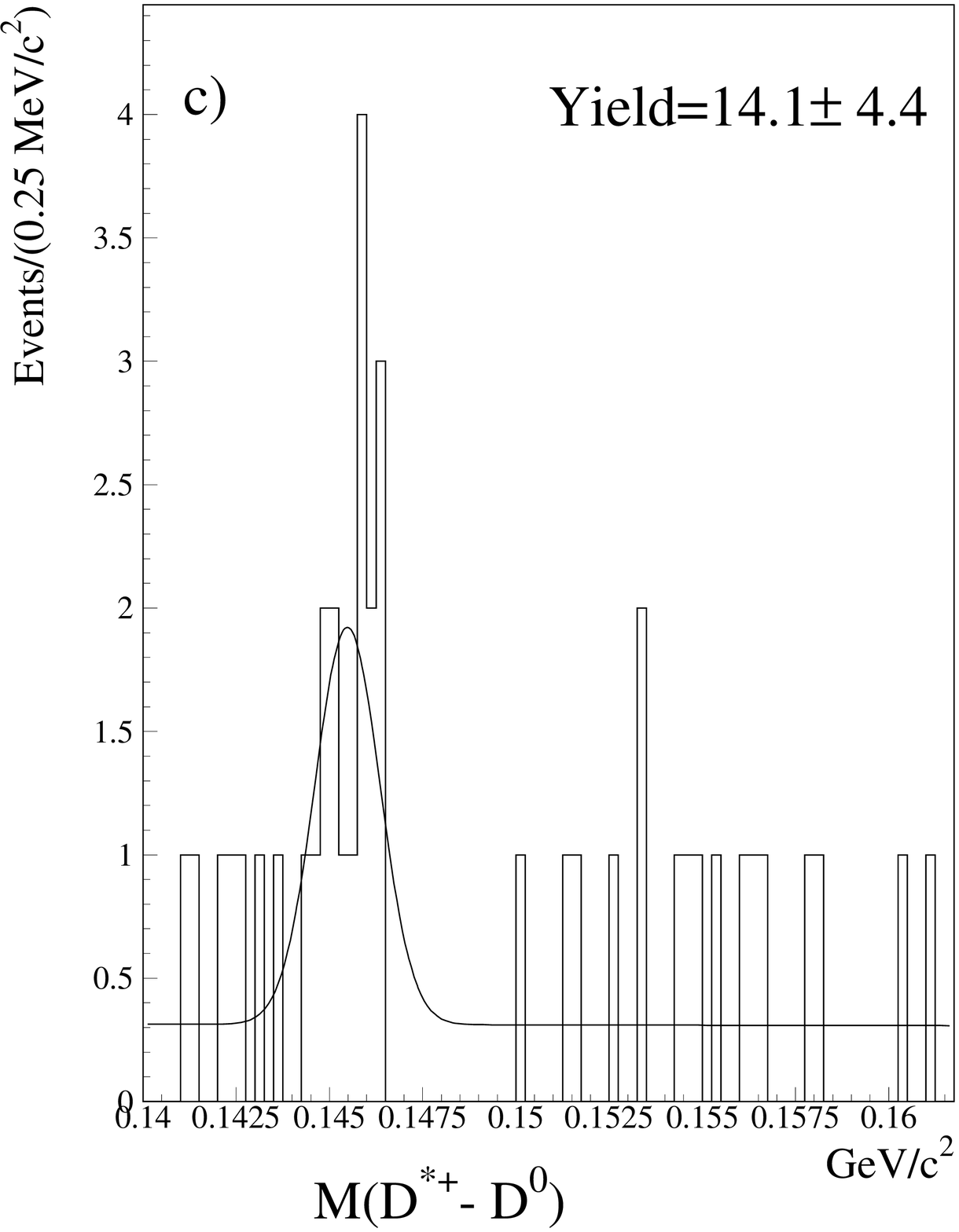}
\end{center}
\caption{Mass difference between 
$M(K_S^0K_S^0K^\pm\pi^\mp\pi_s)$ and $M(K_S^0K_S^0K^\pm\pi^\mp)$ 
where $\pi_s$ is a pion from the primary vertex.  Plots a, b, and c are 
for cases where the $\pi_s$ charge relative to the kaon charge is 
opposite, same, or either, respectively.  A mass cut of $1.85 < 
M(K_S^0K_S^0K^\pm\pi^\mp\pi_s) < 1.88$ $\textrm{GeV}/c^2$ to select $D^0$ 
events is applied.}
\label{figtwo:tag}
\end{figure}

While we have established that we see a new decay channel, in order to separate
$D^0\rightarrow K^0_SK^0_SK^+\pi^-$ and $D^0\rightarrow K^0_SK^0_SK^-\pi^+$ we need
to tag a $D^0$ by using the soft pion from the $D^{*+} \rightarrow D^0\pi^+$ decay.  In Fig.~\ref{figtwo:tag}c
we present the mass difference $M(K^0_SK^0_SK^{\pm}\pi^{\mp}\pi_s) - M(K^0_SK^0_SK^{\pm}\pi^{\mp})$ in
which we find $14.1 \pm 4.4$ signal events when we cut around the $D^0$ mass. Fig.~\ref{figtwo:tag}a~(\ref{figtwo:tag}b) shows the
 mass difference
histogram for the events in which the soft pion has the opposite (same) charge as that of the kaon. 
The two histograms show a signal of $7.2 \pm 3.4$ and $6.8 \pm 2.9$ events, respectively.  The fits in all three histograms
are performed by fixing the mass and width to the values returned from the Monte Carlo.

\section{\bf The $D^0\rightarrow K^0_SK^0_SK^0_S$ Decay Mode}

Although this mode is Cabibbo allowed, it requires either final state interactions or $W$-exchange 
to occur. Due to the limited phase space and the need to identify three $K^0_S$ candidates,  
the signal can be observed without the need for a $D^{*+}$ tag or a $L/\sigma_L$ cut. This significantly
improves the reconstruction efficiency. In 
Fig.~\ref{figone:mass}c  we find a signal of $170 \pm 26$ events. 

We find no evidence for resonant substructure and use a non-resonant
Monte Carlo simulation to obtain the efficiency. 
A $D^0 \rightarrow a_0 (980)^0 K^0_S$ simulation is used as a measure of the
systematic variation (the difference in efficiency is 5.6 \%). For the
branching ratio of this mode with respect to the $D^0
\rightarrow \overline{K}\,\!^0\pi^+\pi^-$ mode we find: 

\begin{equation}
 \frac{\Gamma (D^0 \rightarrow K^0_SK^0_SK^0_S)}
 {\Gamma (D^0 \rightarrow {\overline{K}\,\!^0}\pi^+\pi^-)} = 0.0179 \pm 0.0027 \pm 0.0026.
\end{equation}
We have included systematic uncertainty contributions from split samples (0), fit variations (0.0007),
possible resonant decay (0.0005), Monte Carlo statistics (0.0001), absolute tracking efficiency
(0.0001), and the 
variation between $K_S^0$ reconstruction categories (0.0025).
Several groups \cite{ALBRECHT-3KS,AMMAR-3KS,FRAB-3KS,ASNER-3KS} have reported measurements of
$D^0 \rightarrow K^0_SK^0_SK^0_S$.

\section{\bf The $D^0\rightarrow K^0_SK^0_S$ Decay Mode}

The  $D^0 \rightarrow K^0 \overline{K}\,\!^0$ decay is
 expected to occur primarily via two $W$-exchange diagrams, which are expected to cancel out in a
four-quark model. In the standard six-quark model the difference between the two amplitudes is tiny, proportional
to $SU(3)$ flavour breaking effects, and we thus might expect the branching
fraction for the decay to be very small ($< 3 \times 10^{-4}$ ). However, a
Standard Model based calculation \cite{PHAM-2KS} predicts a relatively large branching fraction
due to final-state interaction effects, leading to a branching ratio of $B(
D^0 \rightarrow K^0 \overline{K}\,\!^0)=\frac{1}{2} B(D^0 \rightarrow K^+ K^-)=0.3$\%.
A recent investigation \cite{DAI-2KS} has focused on the 
$s$-channel and the $t$-channel one particle exchange (OPE) contributions. While the $s$-channel
contribution,
taken into account through the poorly known scalar meson $f_0 (1710)$, gives a small contribution,
the one particle $t$-exchange gives higher contributions, with pion exchange being the highest.
In the factorization limit \cite{BAUER-2KS}, the branching fraction is expected
to be equal to zero. Non-factorizable contributions in factorization-type models have been recently
studied \cite{EEG-2KS} and predict a branching fraction of about 10$^{-4}$.
Several experiments \cite{ASNER-2KS,FRAB-2KS,ALEX-2KS,CUM-2KS} have reported measurements
of $D^0
\rightarrow K^0 _S K^0 _S$.

This channel is similar to $D^0 \rightarrow K^0 _S K^0 _S K^0 _S$ in that no $L/\sigma_L$ cut is
possible, but
the channel requires an additional $D^*$ tag requirement.
We select $D^{*+}$ candidates by requiring that the reconstructed mass difference
$\Delta M=M_{D^{*+}} -M_{D^0}$ lies within 2 MeV/$c^2$ of the nominal mass
difference \cite{PDG} of 145.42 MeV/$c^2$.

The dominant background sources for $D^0\rightarrow K_S^0K_S^0$ decays 
are from non-resonant $K_S^0\pi^+\pi^-$ and $\pi^+\pi^-\pi^+\pi^-$ decays. To reduce feedthrough
from $D^0 \rightarrow K^0 _S \pi^+ \pi^- $ where the $\pi^+ \pi^- $ invariant mass falls in  $K^0
_S$ mass region or from $D^0 \rightarrow \pi^+ \pi^- \pi^+ \pi^- $ where both $\pi^+ \pi^-$ pairs can be misidentified as 
$K ^0 _S$ candidates, we make the additional requirement that the $K^0 _S$ candidates reconstructed with silicon strip
informations have a decay length
significance ($L/\sigma_L$) greater than 12.
We also apply a $|\cos \theta_{K^0 _S}| <$ 0.8  cut, where $\theta_{K^0 _S}$
is the angle between the $K^0 _S$ momentum in the $D^0$ rest frame and the $D^0$
laboratory momentum.

In Fig.~\ref{figthree:mass}a the $K^0 _S K^0 _S$ invariant mass distribution is shown 
for the events satisfying these cuts.
For the signal shape we use a double Gaussian (two Gaussians with the same mean) for the $D^0$ signal 
($79 \pm 17$ events), as suggested by Monte 
Carlo studies, and a first order
Chebyshev polynomial for the background. The double Gaussian shape is fixed to the one obtained from
Monte Carlo simulation.
In Fig.~\ref{figthree:mass}b the $\Delta M = D^{*+} - D^0$ mass difference distribution in the $D^0$ mass
region is plotted.
 A Gaussian
is used to fit the signal while for the background we use the functional form
\begin{equation}
\centering
a (\Delta M- m_{\pi})^{1/2} +b (\Delta M - m_{\pi})^{3/2}
\end{equation}
where $m_{\pi}$ is the pion mass, the first term is from a non--relativistic model of phase space and
the second
term is the first--order relativistic correction to the non-relativistic model.
Consistent yields are obtained from the two fits.

The events of the normalization mode $D^0 \rightarrow K^0_S \pi^+ \pi^- $ have been selected using the
same reconstruction and fitting techniques as for $D^0 \rightarrow K^0 _S K^0_S$ mode and similar selection criteria 
 to minimize systematic uncertainties.

We measure
\begin{equation}
\frac{\Gamma\left(D^0 \rightarrow K_S^0 K_S^0 \right)}{\Gamma\left(D^0 
\rightarrow K_S^0\pi^+\pi^-\right)} =\frac{ \Gamma(D^0 \rightarrow K^0 \overline{K}\,\!^0 )}
{\Gamma(D^0 \rightarrow \overline{K}\,\!^0 \pi^+ \pi^- )}= 0.0144 \pm 0.0032 \pm 0.0016.
\end{equation}

\noindent This accounts for the unseen $K_L^0\pi^+\pi^-$ and $K_L^0K_L^0$ decays 
(note that $D^0 \rightarrow K_S^0K_L^0$ is forbidden by CP conservation).
Different systematic sources have been investigated and the value reported includes contributions from
fit
variations (0.0013) and $K^0 _S$ reconstruction (0.0010).

\section{\bf The $D^0\rightarrow K^0_SK^0_S\pi^+\pi^-$ Decay Mode}

\begin{figure}[t!]
\begin{center}
\includegraphics[width=4.4cm,height=6.8cm]{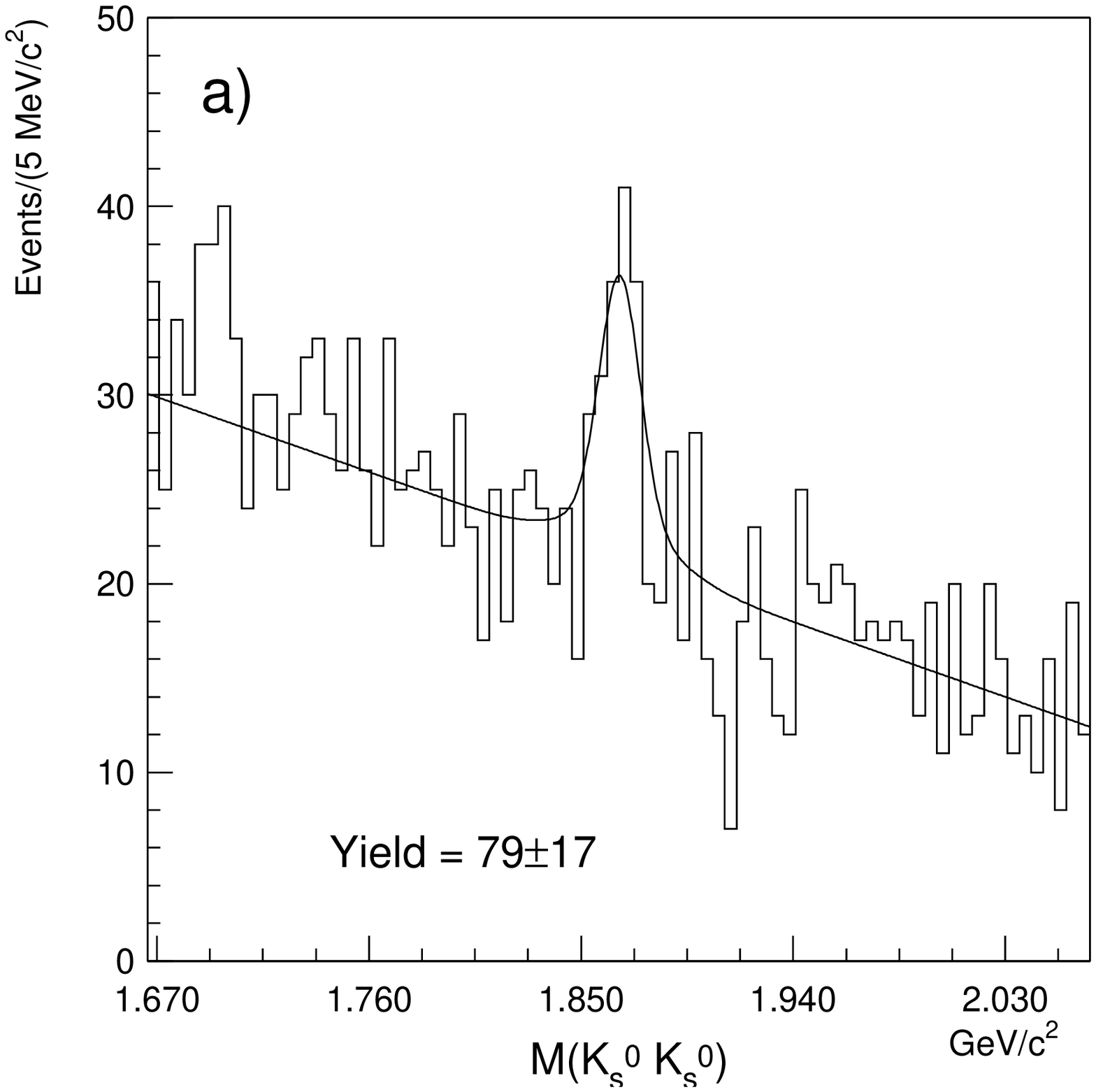}
\includegraphics[width=4.4cm,height=6.8cm]{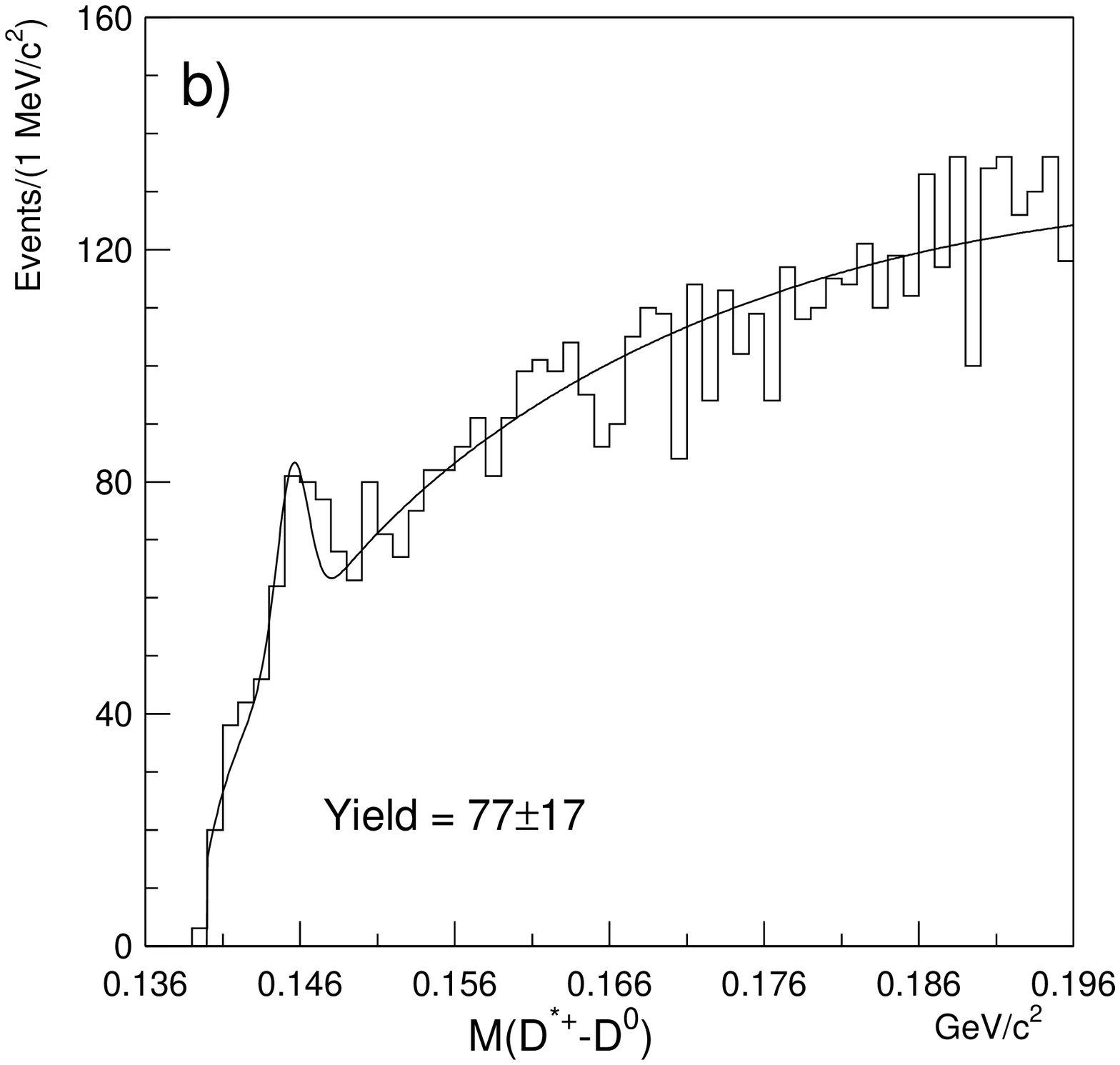}
\includegraphics[width=4.86cm,height=6.9cm]{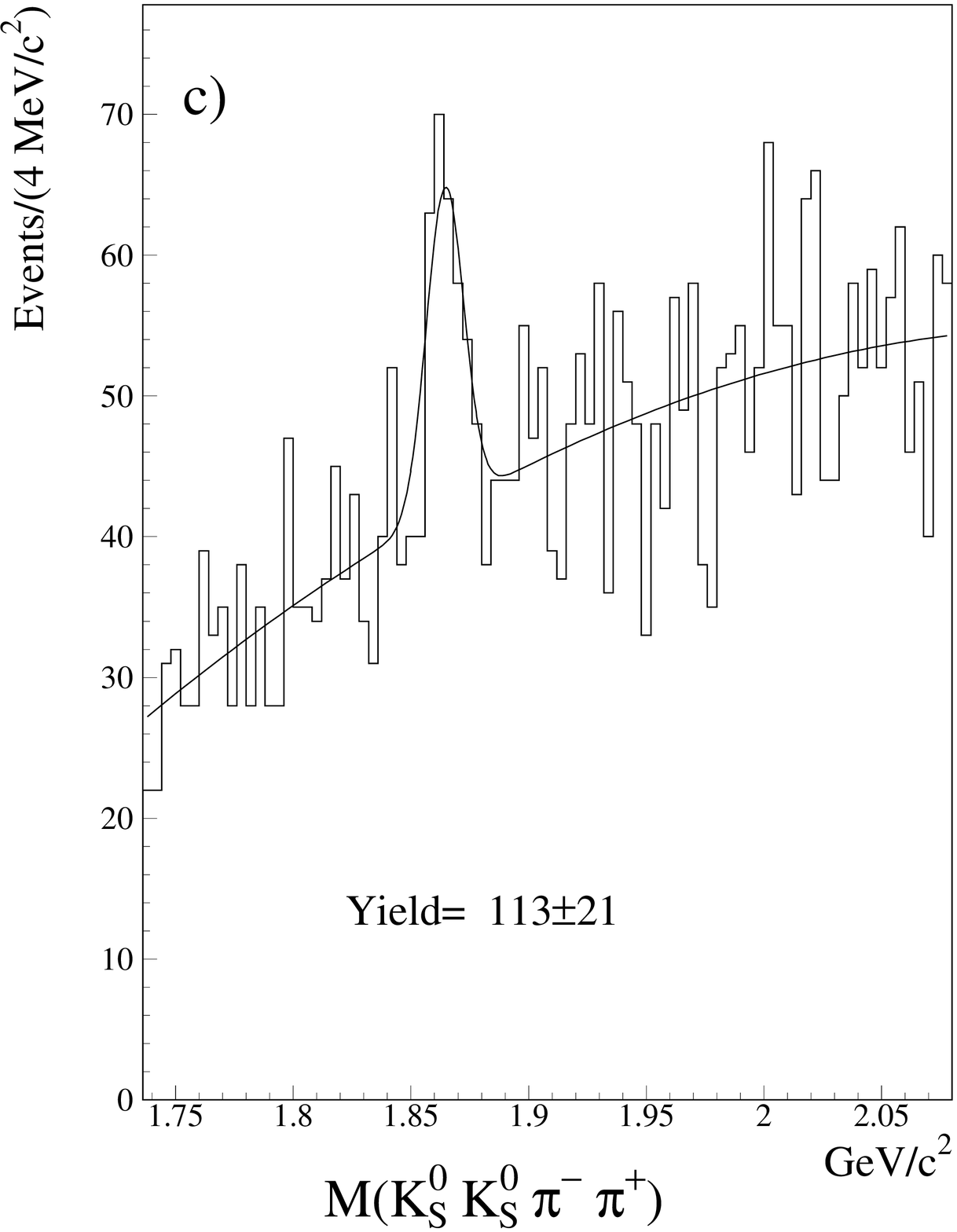}
\end{center}
  \caption{ Invariant mass distribution for various $D^0$ final states:
    (a) Reconstructed mass of $D^0 \rightarrow K^0_SK^0_S$ with a $D^{*+}-D^0$ mass difference
cut. There are $79 \pm 17$ events
with a sigma of 12.5 MeV$/c^2$.
    (b) Reconstructed mass difference of $D^{*+}-D^0; D^0\rightarrow K^0_SK^0_S$. There are $77 \pm 17$
events in the mass difference plot demonstrating consistency with (a).
    (c) Reconstructed mass of $D^0 \rightarrow K^0_SK^0_S\pi^+\pi^-$. There are
    $113 \pm 21$ events.
   }
\label{figthree:mass}
\end{figure}

This channel is Cabibbo-suppressed and was first observed by the ARGUS Collaboration \cite{ALBRECHT-KSKSPIPI}.
It has a larger background than 
the $D^0\rightarrow K^0_SK^0_SK^+\pi^-$ channel
due to more phase space and two pions.  In addition one must eliminate $\pi^+\pi^-$ 
combinations which are consistent with the $K^0_S$ mass. To increase the signal to noise, 
a $L/\sigma > 6$ cut is applied. In Fig.~\ref{figthree:mass}c we present the $K^0_SK^0_S\pi^+\pi^-$
invariant mass. The figure is fit with a Gaussian for the $D^0$ signal plus a $2^{nd}$ order
polynomial. 

As the resonant structure for this channel is not known, we use a non-resonant Monte Carlo
simulation to compute the branching fraction. \\ A $D^0\rightarrow K^*(892)^- K^*(892)^+$ simulation
is used as a measure of the systematic variation (the difference in efficiency is 7\%).  
Again, we calculate the branching ratio of this mode with respect to the $D^0
\rightarrow \overline{K}\,\!^0\pi^+\pi^-$ mode. 
\begin{equation}
 \frac{\Gamma (D^0 \rightarrow K^0_SK^0_S\pi^+\pi^-)}
 {\Gamma (D^0 \rightarrow {\overline{K}\,\!^0}\pi^+\pi^-)} = 0.0208 \pm 0.0035 \pm 0.0021
\end{equation}
We have included systematic uncertainty contributions from split samples (0), fit variations (0.0013),
possible resonant decay (0.0005), Monte Carlo statistics (0.0002), absolute tracking efficiency
(0.0001), and the
variation between $K_S^0$ reconstruction categories (0.0015).

\section{Conclusions}

We summarize our branching ratios and compare them in Table I to the Particle Data Group
world averages where available \cite{PDG}.

\begin{table}[t!]
\begin{center}
\caption{ $D^0 \rightarrow K^0_SK^0_S X$ Branching Fractions}
\vspace{0.15cm}
\begin{tabular}{ccc}
\hline\hline
\rule[-0.3cm]{0pt}{0.8cm}
Decay Mode & This Experiment  & PDG 2004 \cite{PDG} \\
\hline
\rule[-0.3cm]{0pt}{0.5cm}
$\frac{\Gamma(D^0\rightarrow K^0_SK^0_SK^{\pm}\pi^{\mp})}{\Gamma(D^0\rightarrow
\overline{K}\,\!^0\pi^+\pi^-)}$&0.0106 $\pm$ 0.0019 $\pm$ 0.0010& - \\
$\frac{\Gamma(D^0\rightarrow K^0_SK^0_SK^0_S)}{\Gamma(D^0\rightarrow
\overline{K}\,\!^0\pi^+\pi^-)}$&0.0179 $\pm$ 0.0027 $\pm$ 0.0026 & 0.0154 $\pm$ 0.0025 \\
$\frac{\Gamma(D^0\rightarrow K^0\overline{K}\,\!^0)}{\Gamma(D^0\rightarrow
\overline{K}\,\!^0\pi^+\pi^-)}$&0.0144 $\pm$ 0.0032 $\pm$ 0.0016 & 0.0119 $\pm$ 0.0033 \\
$\frac{\Gamma(D^0\rightarrow K^0_SK^0_S\pi^+\pi^-)}{\Gamma(D^0\rightarrow
\overline{K}\,\!^0\pi^+\pi^-)}$&0.0208 $\pm$ 0.0035 $\pm$ 0.0021& 0.031 $\pm$ 0.010 $\pm$ 0.008\\
\hline \hline
\end{tabular}
\label{tb:compare}
\end{center}
\end{table}

We have investigated and measured the branching ratios of several decay modes of the charm meson $D^0$  into
final states containing at least two $K_S^0$ relative to $D^0\rightarrow
\overline{K}\,\!^0\pi^+\pi^-$ mode.
We have evidence for two new Cabibbo favored $D^0$ decay modes, namely
$D^0\rightarrow K^0_SK^0_SK^+\pi^-$ and $D^0\rightarrow K^0_SK^0_SK^-\pi^+$ and we have
reported new measurements for $D^0\rightarrow K^0_SK^0_SK^0_S$, $D^0\rightarrow K^0_SK^0_S$,
and $D^0\rightarrow K^0_SK^0_S\pi^+\pi^-$. 

\input{ack.tex}

\input{bib.tex}
\end{document}

%% file: author.tex
The FOCUS Collaboration\footnote{See http://www-focus.fnal.gov/authors.html for
additional author information.}
\author[ucd]{J.~M.~Link},
\author[ucd]{P.~M.~Yager},
\author[cbpf]{J.~C.~Anjos},
\author[cbpf]{I.~Bediaga},
\author[cbpf]{C.~G\"obel},
\author[cbpf]{A.~A.~Machado},
\author[cbpf]{J.~Magnin},
\author[cbpf]{A.~Massafferri},
\author[cbpf]{J.~M.~de~Miranda},
\author[cbpf]{I.~M.~Pepe},
% ...
 
\address[ucd]{University of California, Davis, CA 95616}
\address[cbpf]{Centro Brasileiro de Pesquisas F\'isicas, Rio de Janeiro, RJ, Brasil}
\address[cinv]{CINVESTAV, 07000 M\'exico City, DF, Mexico}
\address[cu]{University of Colorado, Boulder, CO 80309}
\nopagebreak
\address[fnal]{Fermi National Accelerator Laboratory, Batavia, IL 60510}
\address[fras]{Laboratori Nazionali di Frascati dell'INFN, Frascati, Italy I-00044}
\address[mex]{University of Guanajuato, 37150 Leon, Guanajuato, Mexico}
\address[ui]{University of Illinois, Urbana-Champaign, IL 61801}
\address[iu]{Indiana University, Bloomington, IN 47405}
\address[korea]{Korea University, Seoul, Korea 136-701}
\address[korea2]{Kyungpook National University, Taegu, Korea 702-701}
\address[milan]{INFN and University of Milano, Milano, Italy}
\address[nc]{University of North Carolina, Asheville, NC 28804}
\address[pavia]{Dipartimento di Fisica Nucleare e Teorica and INFN, Pavia, Italy}
\address[pr]{University of Puerto Rico, Mayaguez, PR 00681}
\address[sc]{University of South Carolina, Columbia, SC 29208}
\address[ut]{University of Tennessee, Knoxville, TN 37996}
\address[vu]{Vanderbilt University, Nashville, TN 37235}
\address[wisc]{University of Wisconsin, Madison, WI 53706}

%% file: ack.tex
\section{\label{sec:level8}Acknowledgments} 
~
We wish to acknowledge the assistance of the staffs of Fermi National
Accelerator Laboratory, the INFN of Italy, and the physics departments
of
the
collaborating institutions. This research was supported in part by the
U.~S.
National Science Foundation, the U.~S. Department of Energy, the Italian
Istituto Nazionale di Fisica Nucleare and 
Ministero della Istruzione Universit\`a e
Ricerca, the Brazilian Conselho Nacional de
Desenvolvimento Cient\'{\i}fico e Tecnol\'ogico, CONACyT-M\'exico, and
the Korea Research Foundation of the  
Korean Ministry of Education.